%% file: ms_2.tex
\documentclass{elsart}

\def  \betabf  {{\mbox{\boldmath$\beta$}}}
\def  \gammabf  {{\mbox{\boldmath$\gamma$}}}
\def  \gammabfscript  {{\mbox{\scriptsize\boldmath$\gamma$}}}

\usepackage{natbib}
\usepackage{lscape}
\usepackage{amssymb}

\journal{Accident Analysis and Prevention}

\begin{document}

\begin{frontmatter}


\title{Zero-state Markov switching count-data models:
an empirical assessment}

\author{Nataliya V. Malyshkina\corauthref{correspond_author}} and
\corauth[correspond_author]{Corresponding author.}
\ead{nmalyshk@purdue.edu}
\author{Fred L. Mannering}
\ead{flm@ecn.purdue.edu}
\address{School of Civil Engineering, 550 Stadium Mall Drive,
Purdue University, West Lafayette, IN 47907, United States}

\begin{abstract}
In this study, a two-state Markov switching count-data model is
proposed as an alternative to zero-inflated models to account
for the preponderance of zeros sometimes observed in transportation
count data, such as the number of accidents occurring on a roadway
segment over some period of time. For this accident-frequency case,
zero-inflated models assume the existence of two states:
one of the states is a zero-accident count state, in which 
accident probabilities are so low that they cannot be
statistically distinguished from zero,
and the other state is a normal count state, in which counts can be
non-negative integers that are generated by some counting
process, for example, a Poisson or negative binomial. In contrast
to zero-inflated models, Markov switching models allow specific
roadway segments to switch between the two states over time. An
important advantage of this Markov switching approach is that it
allows for the direct statistical estimation of the specific
roadway-segment state (i.e., zero or count state) whereas traditional
zero-inflated models do not. To demonstrate the applicability of
this approach, a two-state Markov switching negative binomial model
(estimated with Bayesian inference) and standard zero-inflated
negative binomial models are estimated using five-year accident
frequencies on Indiana interstate highway segments. It is shown that
the Markov switching model is a viable alternative and results in a
superior statistical fit relative to the zero-inflated models.
\end{abstract}


\end{frontmatter}

\section{Introduction}
\label{INTRO}

The preponderance of zeros observed in many count-data applications
has lead researchers to consider the possibility that two states
exist; one state that is a ``zero'' state (where all counts are zero)
and the other that is a normal count state that includes zeros and
positive integers. This two-state assumption has led to the
development of zero-inflated Poisson models and zero-inflated
negative binomial models to account for possible overdispersion
in the normal-count state. These zero-inflated models have been
applied to a number of fields of study.  For example, \citet[][]{L_92}
used a zero-inflated Poisson model to study manufacturing defects.
Lambert argued that unobserved changes in the process caused
manufacturing defects to move randomly between a state that was
near perfect (the zero state where defects were extremely rare) and
an imperfect state where defects were possible but not inevitable
(the normal count state). Lambert's empirical assessment
demonstrated that the zero-inflated modeling approach fit the
data much better than the standard Poisson. In other work, van den
\citet[][]{B_95} provided an application of the zero-inflated
Poisson to the frequency of urinary tract infections in men
diagnosed with the human immunodeficiency virus (HIV). In this
case, it was postulated that a zero-infection state existed for
a portion of the patient population and that this state generated
a large number of zeros in the frequency data, which was supported
by the statistical findings. Also, \citet[][]{BDSMK_99} successfully
applied the zero-inflated Poisson to study the frequency of dental
decay in Portugal.

The frequency of vehicle accidents on a section of highway or at an
intersection (over some time period) often exhibit excess zeros.
Similar to the literature discussed above, the excess of zeros
observed in the data could potentially be explained by the existence
of a two-state process for accident data generation
\citep[][]{SMM_97,CM_01,LM_02}. In this case, roadway segments can
belong to one of two states: a zero-accident state (where zero accidents
are expected) and a normal-count state, in which accidents can happen
and accident frequencies are generated by some given counting
process (Poisson or negative binomial). To account for the two-state
phenomena, zero-inflated Poisson (ZIP) and zero-inflated negative
binomial (ZINB) models have been used in a number of roadway safety
studies \citep[][]{M_94,SMM_97,WKM_03}. These models explicitly
account for an existence of the two states for accident data
generation and allow modeling of the probabilities of being in these
states.

An application of ZIP and ZINB models was an empirical advance in
statistical modeling of accident frequencies. However, although
zero-inflated models have become popular in a number of fields, they
suffer from two important drawbacks. First, these models do not deal
directly with the states of roadway segments, instead they consider
probabilities of being in these states. As a result, zero-inflated
models do not allow a direct statistical estimation of whether
individual roadway segments are in the zero or normal count state.
For example, suppose a given roadway segment has zero accidents
observed over a given time interval. This segment could truly be in
the zero-accident count state, or it may be in the normal-count 
state and just happened to have zero accidents over the
considered time interval \citep[][]{SMM_97}. Distinguishing between
these two possibilities is not straightforward in zero-inflated
models. The second drawback of zero-inflated models is that, although
they allow roadway segments to be in different states during different
observation periods, zero-inflated models do not explicitly consider
switching by the roadway segments between the states over time. This
switching is important from the theoretical point of view because it
is unreasonable to expect any roadway segment to be in the zero-accident 
all the time and to have the long-term mean accident frequency equal 
to zero \citep[][]{LWI_05}.

In this study, we propose two-state Markov switching count-data models
that consider the zero-accident state and the normal-count state of roadway
safety. Similar to zero-inflated models, Markov switching models are
intended to explain the preponderance of zeros observed in accident
count data. However, in contrast to zero-inflated models, Markov
switching models allow a direct statistical estimation of the states
roadway segments are in at specific points in time and explicitly
consider changes in these states over time.

\section{Model specification}
\label{MOD_SPECIF}

Two-state Markov switching count-data models of accident frequencies
were first presented in \citet[][]{MMT_09}. Following that paper,
we note that, although there are several major differences
between \citet[][]{MMT_09} and this study, many ideas and
statistical estimation methods developed in \citet[][]{MMT_09}
apply in this study as well. In that paper, two states were assumed
to exist but both were true count states (i.e., a zero-count state
did not exist).  In the current paper, we take a different approach
and consider the case where one of the states is a zero state and
the other is a true count state and that individual roadway segments
move between these two states over time. This differs from
\citet[][]{MMT_09} in that their model assumes two true-count states
and that all roadway segments are in the same state at the same time.

To show this model, we note that Markov switching models are
parametric and can be fully specified by a likelihood function
$f({\bf Y}|{\bf\Theta},{\cal M})$, which is the conditional probability
distribution of the vector of all observations ${\bf Y}$, given the
vector of all parameters ${\bf\Theta}$ of model ${\cal M}$. In our study,
we observe the number of accidents $A_{t,n}$ that occur on the $n^{\rm th}$
roadway segment during time period $t$. Thus ${\bf Y}=\{A_{t,n}\}$
includes all accidents observed on all roadway segments over all
time periods. Here $n=1,2,\ldots,N$ and $t=1,2,\ldots,T$, where $N$ is
the total number of roadway segments observed (it is assumed to be
constant over time) and $T$ is the total number of time periods. Model
${\cal M}=\{M,{\bf X}_{t,n}\}$ includes the model's name $M$ (for example,
$M=\mbox{``ZIP'' or ``ZINB''}$) and the vector ${\bf X}_{t,n}$ of all
roadway segment characteristic variables (segment length, curve
characteristics, grades, pavement properties, and so on).

To define the likelihood function, we introduce an unobserved (latent)
state variable $s_{t,n}$, which determines the state of the $n^{\rm th}$
roadway segment during time period $t$. Without loss of generality, it
is assumed assume that the state variable $s_{t,n}$ can take on the
following two values: $s_{t,n}=0$ corresponds to the zero-accident
state, and $s_{t,n}=1$ corresponds to the normal-count state ($n=1,2,\ldots,N$
and $t=1,2,\ldots,T$). It is further assumed that, for each roadway
segment $n$, the state variable $s_{t,n}$ follows a stationary two-state
Markov chain process in time,\footnote{\label{MARKOV}Markov property
means that the probability distribution of $s_{t+1,n}$ depends only on
the value $s_{t,n}$ at time $t$, but not on the previous history
$s_{t-1},s_{t-2},\ldots$. Stationarity of $\{s_{t,n}\}$ is in the
statistical sense.} which can be specified by time-independent transition
probabilities as
\begin{eqnarray}
P(s_{t+1,n}=1|s_{t,n}=0)=p^{(n)}_{0\rightarrow1},
\quad P(s_{t+1,n}=0|s_{t,n}=1)=p^{(n)}_{1\rightarrow0}.
\label{EQ_P}
\end{eqnarray}
Here, for example, $P(s_{t+1,n}=1|s_{t,n}=0)$ is the conditional
probability of $s_{t+1,n}=1$ at time $t+1$, given that $s_{t,n}=0$
at time $t$. Transition probabilities $p^{(n)}_{0\rightarrow1}$ and
$p^{(n)}_{1\rightarrow0}$ are unknown parameters to be estimated from
accident data ($n=1,2,\ldots,N$). The stationary unconditional
probabilities of states $s_{t,n}=0$ and $s_{t,n}=1$ are
$\bar p^{(n)}_0=p^{(n)}_{1\rightarrow0}/(p^{(n)}_{0\rightarrow1}
+p^{(n)}_{1\rightarrow0})$ and
$\bar p^{(n)}_1=p^{(n)}_{0\rightarrow1}/(p^{(n)}_{0\rightarrow1}
+p^{(n)}_{1\rightarrow0})$ respectively.\footnote{These
can be found from stationarity conditions
$\bar p^{(n)}_0=[1-p^{(n)}_{0\rightarrow1}]\bar p^{(n)}_0
+p^{(n)}_{1\rightarrow0}\bar p^{(n)}_1$,
$\bar p^{(n)}_1=p^{(n)}_{0\rightarrow1}\bar p^{(n)}_0
+[1-p^{(n)}_{1\rightarrow0}]\bar p^{(n)}_1$
and $\bar p^{(n)}_0+\bar p^{(n)}_1=1$.}
If $p^{(n)}_{0\rightarrow1}<p^{(n)}_{1\rightarrow0}$, then $\bar
p^{(n)}_{0}>\bar p^{(n)}_{1}$ and, on average, for roadway
segment $n$ state $s_{t,n}=0$ occurs more frequently than
state $s_{t,n}=1$. If
$p^{(n)}_{0\rightarrow1}>p^{(n)}_{1\rightarrow0}$, then state
$s_{t,n}=1$ occurs more frequently for segment $n$.\footnote{Here,
Eq.~(\ref{EQ_P}) is a significant departure from \citet[][]{MMT_09}
in that individual roadway segments can be in different states
at the same time (i.e., the state variable is subscripted by
roadway segment $n$). Also, in contrast to \citet[][]{MMT_09}, here
we do not restrict state $s_{t,n}=0$ to be more frequent than
state $s_{t,n}=1$.}

Next, consider a two-state Markov switching
negative binomial (MSNB) model that assumes a negative binomial (NB)
data-generating process in the normal-count state $s_{t,n}=1$. With this,
the probability of $A_{t,n}$ accidents occurring on roadway
segment $n$ during time period $t$ is
\begin{eqnarray}
&& P_{t,n}^{(A)} = \left\{
    \begin{array}{ll}
    {\cal I}(A_{t,n}) & \quad\mbox{if $s_{t,n}=0$}\\
    {\cal NB}(A_{t,n})  & \quad\mbox{if $s_{t,n}=1$}
    \end{array}
\right.,
\label{EQ_L_MSNB}\\
&& {\cal I}(A_{t,n}) = \left\{
\mbox{$\,1$ if $A_{t,n}=0$ and $0$ if $A_{t,n}>0\,$}\right\},
\label{EQ_I}\\
&& {\cal NB}(A_{t,n})=\frac{\Gamma(A_{t,n}+1/\alpha)}
{\Gamma(1/\alpha)A_{t,n}!}
\left(\frac{1}{1+\alpha\lambda_{t,n}}\right)^{1/\alpha}
\left(\frac{\alpha\lambda_{t,n}}{1+\alpha\lambda_{t,n}}\right)^{A_{t,n}},
\label{EQ_NB}\\
&& \lambda_{t,n}=\exp(\betabf'{\bf X}_{t,n}),
\quad t=1,2,\ldots,T,\quad n=1,2,\ldots,N. \vphantom{\int}
\label{EQ_LAMBDA}
\end{eqnarray}
Here, Eq.~(\ref{EQ_I}) is the probability mass function that
reflects the fact that accidents never happen in the zero-accident
state $s_{t,n}=0$.\footnote{Although Eq.~(\ref{EQ_I}) formally assumes
$s_{t,n}=0$ to be a zero-accident state, in which accidents never happen,
this state can be viewed as an approximation for a nearly safe state,
in which the average accident rate is negligible ($\lambda_{t,n}\ll 1$)
and accidents are extremely rare (over the considered time period).
}
Eq.~(\ref{EQ_NB}) is the standard negative
binomial probability mass function, $\Gamma(\:)$ is the gamma
function, and prime means transpose (so $\betabf'$ is the
transpose of $\betabf$). Parameter vector $\betabf$ and
the over-dispersion parameter $\alpha\ge0$ are unknown estimable
model parameters.\footnote{To ensure that $\alpha$ is non-negative,
we estimate its logarithm instead of it.}
Scalars $\lambda_{t,n}$ are the accident rates in the normal-count state.
We set the first component of
${\bf X}_{t,n}$ to unity, and, therefore, the first component of
$\betabf$ is the intercept.

\begin{figure}[t]
\vspace{7.0truecm}
\includegraphics{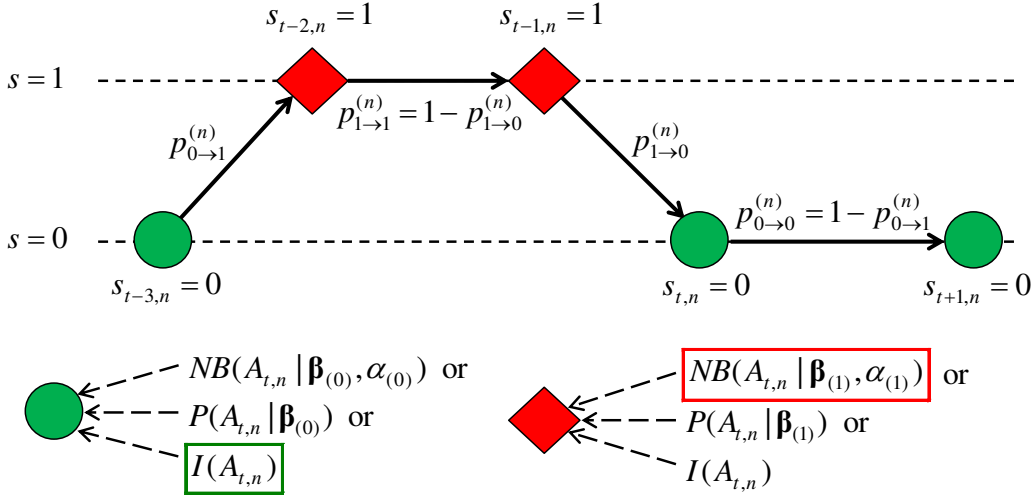}
\caption{Graphical demonstration of a two-state Markov switching model.
\label{FIGURE_1}
}
\end{figure}

A two-state Markov switching model of accident frequencies is
graphically demonstrated in Figure~\ref{FIGURE_1}. In the two 
states $s=0$ and $s=1$ shown in the figure, the
accident frequency data are generated by two different processes,
shown by the circles (for state $s=0$) and the diamonds (for $s=1$).
In this study, we assume that accident frequency is generated according to the
zero-accident distribution ${\cal I}(A_{t,n})$ in state $s=0$, and according to
the standard negative binomial distribution ${\cal NB}(A_{t,n})$ in state $s=1$
(these two distributions are outlined by the boxes in Figure~\ref{FIGURE_1}).
The state variable $s_{t,n}$ follows a Markov process over time, with transition
probabilities $p^{(n)}_{0\rightarrow0}$, $p^{(n)}_{0\rightarrow1}$,
$p^{(n)}_{1\rightarrow0}$ and $p^{(n)}_{1\rightarrow1}$, as shown in
Figure~\ref{FIGURE_1}.

If accident events are assumed to be independent, the
likelihood function is
\begin{eqnarray}
f({\bf Y}|{\bf\Theta},{\cal M})
=\prod\limits_{t=1}^T\prod\limits_{n=1}^N P_{t,n}^{(A)}.
\label{EQ_L}
\end{eqnarray}
Here, because the state variables $s_{t,n}$ are unobservable, the vector
of all estimable parameters ${\bf\Theta}$ must include all states, in
addition to all model parameters ($\beta$-s, $\alpha$) and transition
probabilities. Thus,
${\bf\Theta}=[\betabf',\alpha,p^{(1)}_{0\rightarrow1},\ldots,p^{(N)}_{0\rightarrow1},
\linebreak
p^{(1)}_{1\rightarrow0}, \ldots, p^{(N)}_{1\rightarrow0},{\bf S}']'$,
where vector ${\bf S}=[(s_{1,1},...,s_{T,1}),\ldots,(s_{1,N},...,s_{T,N})]'$
has length $T\times N$ and contains all state values.

Eqs.~(\ref{EQ_P})-(\ref{EQ_L}) define the two-state Markov switching
negative binomial (MSNB) model considered here. Note that in this
model the estimable state variables $s_{t,n}$ explicitly specify the
states of  all roadway segments $n=1,2,\ldots,N$ during all
time periods $t=1,2,\ldots,T$.

In this study, in addition to the MSNB model, we also consider the
standard zero-inflated negative binomial (ZINB) models. In this case,
the probability of $A_{t,n}$ accidents occurring is \citep[][]{WKM_03}
\begin{eqnarray}
P_{t,n}^{(A)} &=& q_{t,n}{\cal I}(A_{t,n})+(1-q_{t,n}){\cal NB}(A_{t,n}),
\label{EQ_L_ZINB}
\\
q_{t,n}&=&\frac{1}{1+e^{-\tau\log\lambda_{t,n}}},
\label{EQ_Q_TAU}
\\
q_{t,n}&=&\frac{1}{1+e^{-\gammabfscript'{\bf X}_{t,n}}},
\label{EQ_Q_GAMMA}
\end{eqnarray}
where we use two different specifications for the probability
$q_{t,n}$ that the $n^{\rm th}$ roadway segment is in the
zero-accident state during time period $t$. The right-hand-side
of Eq.~(\ref{EQ_L_ZINB}) is a mixture of zero-accident
distribution ${\cal I}(A_{t,n})$ given by Eq.~(\ref{EQ_I}) and
negative binomial distribution ${\cal NB}(A_{t,n})$ given
by Eq.~(\ref{EQ_NB}). Scalar $\tau$ and vector $\gammabf$ are
estimable model parameters. Accident rate $\lambda_{t,n}$ is given
by Eq.~(\ref{EQ_LAMBDA}). We call ``\mbox{ZINB-$\tau$}'' the
model specified by Eqs.~(\ref{EQ_L_ZINB}) and~(\ref{EQ_Q_TAU}).
We call ``\mbox{ZINB-$\gamma$}'' the model specified by
Eqs.~(\ref{EQ_L_ZINB}) and~(\ref{EQ_Q_GAMMA}).
Note that $q_{t,n}$ depends on the estimable model parameters
and gives the probability of being in the zero-accident state
$s_{t,n}=0$, but it is not an estimable parameter by itself and
does not explicitly specify the state value $s_{t,n}$.

\section{Model estimation methods}
\label{MOD_ESTIM}

Statistical estimation of Markov switching models is complicated by
unobservability of the state variables
$s_{t,n}$.\footnote{\label{FN_S}Below we will have five time periods
($T=5$) and 335 roadway segments ($N=335$). In this case, there are
$2^{TN}=2^{1675}$ possible combinations for value of vector
${\bf S}=[(s_{1,1},...,s_{T,1}),\ldots,(s_{1,N},...,s_{T,N})]'$.}
As a result, the traditional maximum likelihood estimation (MLE)
procedure is of very limited use for Markov switching models.
Instead, a Bayesian inference approach is used.
Given a model ${\cal M}$ with likelihood function
$f({\bf Y}|{\bf\Theta},{\cal M})$, the Bayes formula is
\begin{eqnarray}
f({\bf\Theta}|{\bf Y},{\cal M})=
\frac{f({\bf Y},{\bf\Theta}|{\cal M})}{f({\bf Y}|{\cal M})}=
\frac{f({\bf Y}|{\bf\Theta},{\cal M})\pi({\bf\Theta}|{\cal M})}
{\int f({\bf Y},{\bf\Theta}|{\cal M})\,d{\bf\Theta}}.
\label{EQ_POSTERIOR}
\end{eqnarray}
Here $f({\bf\Theta}|{\bf Y},{\cal M})$ is the posterior probability
distribution of model parameters ${\bf\Theta}$ conditional on the
observed data ${\bf Y}$ and model ${\cal M}$.
Function $f({\bf Y},{\bf\Theta}|{\cal M})$ is the
joint probability distribution of ${\bf Y}$ and ${\bf\Theta}$ given
model ${\cal M}$. Function $f({\bf Y}|{\cal M})$ is the marginal
likelihood function -- the probability distribution of data
${\bf Y}$ given model ${\cal M}$. Function $\pi({\bf\Theta}|{\cal M})$
is the prior probability distribution of parameters that reflects
prior knowledge about ${\bf\Theta}$. The intuition behind
Eq.~(\ref{EQ_POSTERIOR}) is straightforward: given model ${\cal M}$, the
posterior distribution accounts for both the observations ${\bf Y}$
and our prior knowledge of ${\bf\Theta}$.

In our study (and in most practical studies), the direct application
of Eq.~(\ref{EQ_POSTERIOR}) is not feasible because the parameter vector
${\bf\Theta}$ contains too many components, making integration over
${\bf\Theta}$ in Eq.~(\ref{EQ_POSTERIOR}) extremely difficult. However,
the posterior distribution $f({\bf\Theta}|{\bf Y},{\cal M})$ in
Eq.~(\ref{EQ_POSTERIOR}) is known up to its normalization constant,
$f({\bf\Theta}|{\bf Y},{\cal M})\propto
f({\bf Y}|{\bf\Theta},{\cal M})\pi({\bf\Theta}|{\cal M})$. As
a result, we use Markov Chain Monte Carlo (MCMC) simulations, which
provide a convenient and practical computational methodology for
sampling from a probability distribution known up to a constant (the
posterior distribution in our case). Given a large enough posterior
sample of parameter vector ${\bf\Theta}$, any posterior expectation
and variance can be found and Bayesian inference can be readily applied.
A reader interested in details is referred to \citet{M_08}, where
we comprehensively describe our choice of the prior distribution
$\pi({\bf\Theta}|{\cal M})$ and the MCMC simulation
algorithm.\footnote{Our priors for $\alpha$, $\beta$-s,
$p_{0\rightarrow1}$ and $p_{1\rightarrow0}$ are flat or nearly flat,
while the prior for the states ${\bf S}$ reflects the Markov process
property, specified by Eq.~(\ref{EQ_P}).}
We used MATLAB language for programming and running the MCMC
simulations.

For comparison of different models we use a formal Bayesian
approach. Let there be two models ${\cal M}_1$ and ${\cal M}_2$
with parameter vectors ${\bf\Theta_1}$ and ${\bf\Theta_2}$
respectively. Assuming that we have equal preferences of these
models, their prior probabilities are
$\pi({\cal M}_1)=\pi({\cal M}_2)=1/2$. In this case, the ratio
of the models' posterior probabilities, $P({\cal M}_1|{\bf Y})$
and $P({\cal M}_2|{\bf Y})$, is equal to the Bayes factor. The
later is defined as the ratio of the models' marginal likelihoods
\citep[see][]{KR_95}. Thus, we have
\begin{eqnarray}
\frac{P({\cal M}_2|{\bf Y})}{P({\cal M}_1|{\bf Y})}=
\frac{f({\cal M}_2,{\bf Y})/f(\bf Y)}{f({\cal M}_1,{\bf Y})/f(\bf Y)}=
\frac{f({\bf Y}|{\cal M}_2)\pi({\cal M}_2)}{f({\bf Y}|{\cal M}_1)\pi({\cal M}_1)}=
\frac{f({\bf Y}|{\cal M}_2)}{f({\bf Y}|{\cal M}_1)},
\label{EQ_BAYES_FACTOR}
\end{eqnarray}
where $f({\cal M}_1,{\bf Y})$ and $f({\cal M}_2,{\bf Y})$ are
the joint distributions of the models and the data, $f({\bf Y})$
is the unconditional distribution of the data. As in \citet[][]{MMT_09}, to
calculate the marginal likelihoods $f({\bf Y}|{\cal M}_1)$ and
$f({\bf Y}|{\cal M}_2)$, we use the harmonic mean formula
$f({\bf Y}|{\cal M})^{-1}=
E\left[\left.f({\bf Y}|{\bf\Theta},{\cal M})^{-1}\right|{\bf Y}\right]$,
where $E(\ldots|{\bf Y})$ means posterior expectation calculated
by using the posterior distribution. If the
ratio in Eq.~(\ref{EQ_BAYES_FACTOR}) is larger than one, then model
${\cal M}_2$ is favored, if the ratio is less than one,
then model ${\cal M}_1$ is favored. An advantage of the use of
Bayes factors is that it has an inherent penalty for
including too many parameters in the model and guards against
overfitting.

To evaluate the performance of model $\{{\cal M},{\bf\Theta}\}$
in fitting the observed data ${\bf Y}$, we carry out a $\chi^2$
goodness-of-fit test \citep{MS_96,C_98,W_02,PTVF_07}. We perform
this test by Monte Carlo simulations to find the distribution of
the $\chi^2$ quantity, which measures the discrepancy between the
observations and the model predictions \citep{C_98}. This
distribution is then used to find the goodness-of-fit p-value,
which is the probability that $\chi^2$ exceeds the observed value
of $\chi^2$ under the hypothesis that the model is true (the
observed value of $\chi^2$ is calculated by using the observed
data ${\bf Y}$). For additional details, please see
\citet[][]{M_08}.

\section{Empirical results}
\label{RESULTS}

Data are used from 5769 accidents that were
observed on 335 interstate highway segments in
Indiana in 1995-1999. We use annual time periods, $t=1,2,3,4,T=5$ in
total.\footnote{We also considered quarterly time periods and obtained
qualitatively similar results (not reported here).}
Thus, for each roadway segment $n=1,2,\ldots,N=335$ the state
$s_{t,n}$ can change every year. Four types of accident frequency
models are estimated:
\begin{enumerate}
\item First, for the purpose of explanatory variable selection,
we estimate an auxiliary standard negative binomial (NB) model, which
is not reported here. We estimate this model by maximum likelihood
estimation (MLE). To obtain a standard NB model, we choose explanatory
variables and their dummies by using the Akaike Information Criterion
(AIC)\footnote{\label{FN_AIC}Minimization of $AIC=2K-2LL$, were $K$
is the number of free continuous model parameters and $LL$ is the
log-likelihood, ensures an optimal choice of explanatory variables
in a model and avoids overfitting \citep[][]{T_02,WKM_03}.}
and the $5\%$ statistical significance level for the two-tailed
t-test \citep[for details on our variable selection methods,
see][]{M_06}. In order to make a comparison of explanatory variable
effects in different models straightforward, in all other models,
described below, we use only those explanatory variables that enter
the standard NB model.\footnote{A formal Bayesian approach to model
variable selection is based on evaluation of model's marginal
likelihood and the Bayes factor~(\ref{EQ_BAYES_FACTOR}).
Unfortunately, because MCMC simulations are computationally expensive,
evaluation of marginal likelihoods for a large number of trial models
is not feasible in our study.}
\item
We estimate the standard \mbox{ZINB-$\tau$} model, specified by
Eqs.~(\ref{EQ_L})--(\ref{EQ_Q_TAU}). First, we estimate this model
by maximum likelihood estimation (MLE) and use the $5\%$ statistical
significance level for evaluation of the statistical significance
of each $\beta$-parameter. Second, we estimate the
same \mbox{ZINB-$\tau$} model by the Bayesian inference approach
and MCMC simulations. As one expects, the Bayesian-MCMC estimation
results turned out to be similar to the MLE estimation results for
the \mbox{ZINB-$\tau$} model.
\item
We estimate the standard \mbox{ZINB-$\gamma$} model, specified by
Eqs.~(\ref{EQ_L}),~(\ref{EQ_L_ZINB}) and~(\ref{EQ_Q_GAMMA}). First,
we estimate this model by MLE and use the $5\%$ statistical
significance level for evaluation of the statistical significance
of each $\beta$-parameter. Second, we estimate the
same \mbox{ZINB-$\gamma$} model by the Bayesian inference approach
and MCMC simulations. The Bayesian-MCMC and the MLE estimation
results for the \mbox{ZINB-$\gamma$} model turned out to be
similar.
\item
We estimate the two-state Markov switching negative binomial (MSNB)
model, specified by Eqs.~(\ref{EQ_P})-(\ref{EQ_L}), by the
Bayesian-MCMC methods. We consecutively
construct and use $60\%$, $85\%$ and $95\%$ Bayesian credible intervals
for evaluation of the statistical significance of each $\beta$-parameter
in the MSNB model. As a result, in the final MSNB model some components
of $\betabf$ are restricted to zero.\footnote{A $\beta$-parameter is
restricted to zero if it is statistically insignificant. A $\,1-a\,$
credible interval is chosen in such way that the posterior
probabilities of being below and above it are both equal to $a/2$ (we
use significance levels $a=40\%,15\%,5\%$).}
No restriction is imposed on the over-dispersion parameter $\alpha$,
which turns out to be significant anyway.
\end{enumerate}

The model estimation results for accident frequencies are given in
Table~\ref{T_1}. Continuous model parameters, $\beta$-s and $\alpha$,
are given together with their $95\%$ confidence intervals (if MLE)
or $95\%$ credible intervals (if Bayesian-MCMC), refer to the
superscript and subscript numbers adjacent to
parameter estimates in Table~\ref{T_1}.\footnote{Note that MLE
assumes asymptotic normality of the estimates, resulting
in confidence intervals being symmetric around the means (a $95\%$
confidence interval is $\pm1.96$ standard deviations around the mean).
In contrast, Bayesian estimation does not require this assumption, and
posterior distributions of parameters and Bayesian credible intervals
are usually non-symmetric.}
Table~\ref{T_2} gives summary statistics of all roadway segment
characteristic variables ${\bf X}_{t,n}$ (except the intercept).

The estimation results show that the MSNB model is strongly favored
by the empirical data, as compared to the standard ZINB models.
Indeed, from Table~\ref{T_1} we
see that the MSNB model provides considerable, $335.69$ and $263.12$,
improvements of the logarithm of the marginal likelihood of the data
as compared to the \mbox{ZINB-$\tau$} and \mbox{ZINB-$\gamma$}
models.\footnote{We use the harmonic mean formula to calculate
the values and the $95\%$ confidence intervals of the
log-marginal-likelihoods given in Table~\ref{T_1}. The confidence
intervals are calculated by bootstrap simulations. For details,
see \citet[][]{MMT_09} or \citet{M_08}.}
Thus, from Eq.~(\ref{EQ_BAYES_FACTOR}), we find that, given the
accident data, the posterior probability of the MSNB model is
larger than the probabilities of the \mbox{ZINB-$\tau$} and
\mbox{ZINB-$\gamma$} models by $e^{335.69}$ and $e^{263.12}$
respectively.\footnote{There are other frequently used model
comparison criteria, for example, the deviance information
criterion,
${\rm DIC}=2E[D({\bf\Theta})|{\bf Y}]-D(E[{\bf\Theta}|{\bf Y}])$,
where deviance
$D({\bf\Theta})\equiv-2\ln[f({\bf Y}|{\bf\Theta},{\cal M})]$
\citep{R_01}. Models with smaller DIC are favored to models with
larger DIC. We find DIC values $5037.3$, $4891.4$, $4261.5$ for
the \mbox{ZINB-$\tau$}, \mbox{ZINB-$\gamma$} and MSNB models
respectively. This means that the MSNB model is favored over
the standard ZINB models. However, DIC is theoretically based on
the assumption of asymptotic multivariate normality of the
posterior distribution, in which case DIC reduces to AIC
\citep{SBCL_02}. As a result, we prefer to rely on a
mathematically rigorous and formal Bayes factor approach to
model selection, as given by Eq.~(\ref{EQ_BAYES_FACTOR}).
}

Let us now consider the maximum likelihood estimation (MLE)
of the standard \mbox{ZINB-$\tau$} and \mbox{ZINB-$\gamma$} models
and an imaginary MLE estimation of the MSNB model. Referring to
Table~\ref{T_1}, the MLE gave maximum log-likelihood values
$-2502.67$ and $-2426.54$ for the \mbox{ZINB-$\tau$} and
\mbox{ZINB-$\gamma$} models. The maximum log-likelihood value
observed during our MCMC simulations for the MSNB model is equal
to $-2049.45$. An imaginary MLE, at its convergence, would give
MSNB log-likelihood value that would be even larger than this
observed value. Therefore, the MSNB model, if estimated by the
MLE, would provide very large, at least $453.22$ and $377.09$,
improvements in the maximum log-likelihood value over the
\mbox{ZINB-$\tau$} and \mbox{ZINB-$\gamma$} models. These
improvements would come with no increase or a decrease in the
number of free continuous model parameters ($\beta$-s, $\alpha$,
$\tau$, $\gamma$-s) that enter the likelihood function.

\begin{landscape}
\begin{table}[p]
\caption{Estimation results for models of accident frequency
(the superscript and subscript numbers to the right of individual}
{parameter estimates are $95\%$ confidence/credible
intervals -- see text for further explanation)}
\label{T_1}
\begin{scriptsize}
\input{ms_2_Table1a.tex}

\end{scriptsize}
\end{table}
\addtocounter{table}{-1}
\begin{table}[p]
\caption{(Continued)}
\begin{scriptsize}
\input{ms_2_Table1b.tex}

\end{scriptsize}
\end{table}
\begin{table}[p]
\caption{Summary statistics of roadway segment characteristic variables}
\label{T_2}
\begin{scriptsize}
\input{ms_2_Table2.tex}

\end{scriptsize}
\end{table}
\end{landscape}

To evaluate the goodness-of-fit for a model, we use the posterior
(or MLE) estimates of all continuous model parameters ($\beta$-s,
$\alpha$, $p^{(n)}_{0\rightarrow1}$, $p^{(n)}_{1\rightarrow0}$)
and generate $10^4$ artificial data sets under the hypothesis
that the model is true.\footnote{Note that the state values $\bf S$
are generated by using $p^{(n)}_{0\rightarrow1}$ and
$p^{(n)}_{1\rightarrow0}$.}
We find the distribution of $\chi^2$ and calculate the
goodness-of-fit p-value for the observed value of $\chi^2$.
For details, see \citep[][]{MMT_09}. The resulting p-values for
our models are given in Table~\ref{T_1}. For the
\mbox{ZINB-$\gamma$} and MSNB models the p-values are
sufficiently large, around $20\%$, which indicates that these
models fit the data reasonably well. At the same time, for
the \mbox{ZINB-$\tau$} model the goodness-of-fit p-value is
only around $0.5\%$, which indicates a much poorer
fit.~\footnote{It is worth to mention that for the auxiliary
standard negative binomial (NB) model, which we do not report
here, the goodness-of-fit p-value was also very poor,
$\approx0.3\%$. This is an expected result because of a
preponderance of zeros in the data, not accounted for in the
NB model.}

The estimation results also show that the over-dispersion
parameter $\alpha$ is higher for the \mbox{ZINB-$\tau$} and
\mbox{ZINB-$\gamma$} models, as compared to the MSNB model
(refer Table~\ref{T_1}). This suggests that
over-dispersed volatility of accident frequencies, which is often
observed in empirical data, could be in part due to the latent
switching between the states of roadway safety.

\begin{figure}[t]
\vspace{9.2truecm}
\includegraphics{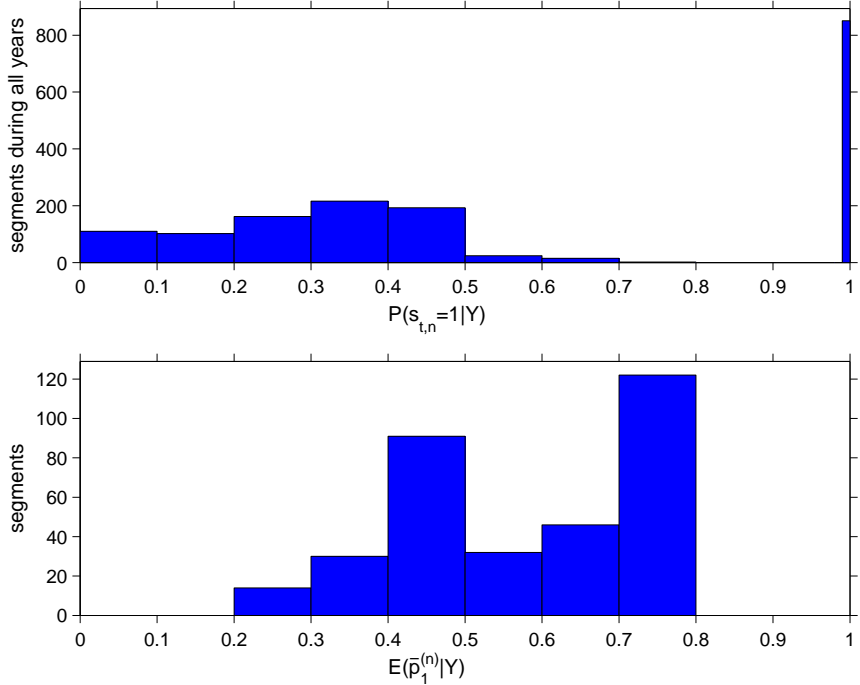}
\caption{Histograms of the posterior probabilities
$P(s_{t,n}=1|{\bf Y})$ (the top plot) and of the posterior
expectations $E[\bar p^{(n)}_1|{\bf Y}]$ (the bottom plot).
Here $t=1,2,3,4,5$ and $n=1,2,\ldots,335$.
\label{FIGURE_3}
}
\end{figure}

Now, refer to Figure~\ref{FIGURE_2}, made for the case of the MSNB
model. The four plots in this figure show five-year time series of
the posterior probabilities $P(s_{t,n}=1|{\bf Y})$ of the normal-count
state for four selected roadway segments.
These plots represent the following four categories of roadway
segments:
\begin{enumerate}
\item
For roadway segments from the first category we have
$P(s_{t,n}=1|{\bf Y})=1$ for all $t=1,2,3,4,5$. Thus, we can say with
absolute certainty that these segments were always in the normal-count state
$s_{t,n}=1$ during the considered five-year time interval. A roadway
segment belongs to this category if and only if it had at least one
accident during each year ($t=1,2,3,4,5$). An example of such roadway
segment is given in the top-left plot in Figure~\ref{FIGURE_2}. For
this segment the posterior expectation of the long-term unconditional
probability ${\bar p}_1$ of being in the normal-count state is large,
$E({\bar p}_1|Y)=0.750$.
\item
For roadway segments from the second category
$P(s_{t,n}=1|{\bf Y})\ll1$ for all $t=1,2,3,4,5$. Thus, we can say
with high degree of certainty that these segments were always in the
zero-accident state $s_{t,n}=0$ during the considered five-year time
interval. A roadway segment $n$ belongs to this category if it had
no accidents observed over the five-year interval despite the
accident rates given by Eq.~(\ref{EQ_LAMBDA}) were large,
$\lambda_{t,n}\gg1$ for all $t=1,2,3,4,5$. Clearly this segment would
be unlikely to have zero accidents observed, if it were not in the
zero-accident state all the time.\footnote{Note that the zero-accident
state may exist due to under-reporting of minor, low-severity accidents
\citep[][]{SMM_97}.}
An example of such roadway segment is given in the top-right plot
in Figure~\ref{FIGURE_2}. For this segment $E({\bar p}_1|Y)=0.260$
is small.

\begin{figure}[t]
\vspace{9.4truecm}
\includegraphics{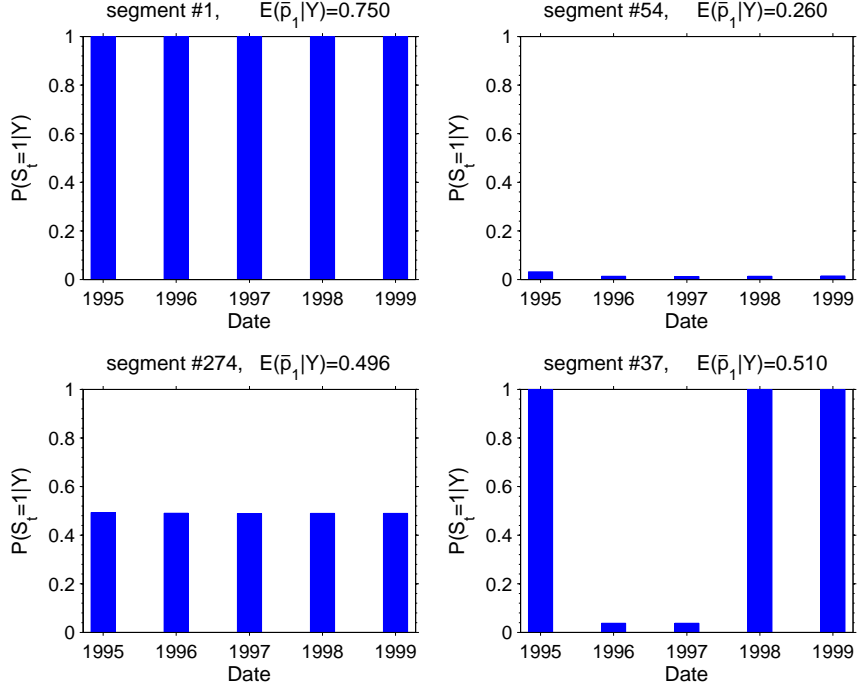}
\caption{Five-year time series of the posterior probabilities
$P(s_{t,n}=1|{\bf Y})$ of the normal-count state $s_{t,n}=1$
for four selected roadway segments ($t=1,2,3,4,5$).
\label{FIGURE_2}
}
\end{figure}

\item
For roadway segments from the third category $P(s_{t,n}=1|{\bf Y})$
is neither one nor close to zero for all $t=1,2,3,4,5$.\footnote{
If there were no Markov switching, which introduces time-dependence
of states via Eqs.~(\ref{EQ_P}), then, assuming non-informative
priors $\pi(s_{t,n}=0)=\pi(s_{t,n}=1)=1/2$ for states $s_{t,n}$,
the posterior probabilities $P(s_{t,n}=1|{\bf Y})$ would be either
exactly equal to $1$ (when $A_{t,n}>0$) or necessarily below
$1/2$ (when $A_{t,n}=0$). In other words, we would have
$P(s_{t,n}=1|{\bf Y})\notin[0.5,1)$ for any $t$ and $n$.
Even with Markov switching existent, in this study we have never
found any $P(s_{t,n}=1|{\bf Y})$ close but not equal to $1$,
refer to the top plot in Figure~\ref{FIGURE_3}.}
For these segments we cannot determine with high certainty what
states these segments were in during years $t=1,2,3,4,5$. A roadway
segment $n$ belongs to this category if it had no any accidents observed
over the considered five-year time interval and the accident rates were
not large, $\lambda_{t,n}\lesssim 1$ for all $t=1,2,3,4,5$. In fact,
when $\lambda_{t,n}\ll1$, the posterior probabilities of the two states
are close to one-half,
$P(s_{t,n}=1|{\bf Y})\approx P(s_{t,n}=0|{\bf Y})\approx0.5$, and
no inference about the value of the state variable $s_{t,n}$ can
be made. In this case of small accident frequencies, the observation of zero
accidents is perfectly consistent with both states $s_{t,n}=0$ and
$s_{t,n}=1$. An example of a roadway segment from the third category
is given in the bottom-left plot in Figure~\ref{FIGURE_2}. For this
segment $E({\bar p}_1|Y)=0.496$ is about one-half.
\item
Finally, the fourth category is a mixture of the three categories
described above. Roadway segments from this fourth category have
posterior probabilities $P(s_{t,n}=1|{\bf Y})$ that change in time
between the three possibilities given above.
In particular, for some roadway segments we can say with high
certainty that they changed their states in time from the
zero-accident state $s_{t,n}=0$ to the normal-count state $s_{t,n}=1$
or vice versa. An example of a roadway segment from the fourth
category is given in the bottom-right plot in Figure~\ref{FIGURE_2}.
For this segment $E({\bar p}_1|Y)=0.510$ is about one-half.
Thus we find a direct empirical evidence that some roadway
segments do change their states over time.
\end{enumerate}

Next, it is useful to consider roadway segment statistics by
state of roadway safety. Referring to Figure~\ref{FIGURE_3},
a case is made for the MSNB model. The top plot in this
figure shows the histogram of the posterior probabilities
$P(s_{t,n}=1|{\bf Y})$ for all $N=335$ roadway segments
during all $T=5$ years ($1675$ values of $s_{t,n}$
in total). For example, we find that during five years roadway
segments had $P(s_{t,n}=1|{\bf Y})=1$ and were normal-count in $851$
cases, and they had $P(s_{t,n}=1|{\bf Y})<0.2$ and were likely
to be zero-accident in $212$ cases. The bottom plot in
Figure~\ref{FIGURE_3} shows the histogram of the posterior
expectations $E[\bar p^{(n)}_1|{\bf Y}]$, where
$\bar p^{(n)}_1=p^{(n)}_{0\rightarrow1}/(p^{(n)}_{0\rightarrow1}
+p^{(n)}_{1\rightarrow0})$
are the stationary unconditional probabilities of
the normal-count state (see Section~\ref{MOD_SPECIF}). We find that
$0.2\le E[\bar p^{(n)}_1|{\bf Y}]\le0.8$ for all segments
$n=1,2,\ldots,335$. This means that in the long run, all
roadway segments have significant probabilities of visiting
both the zero-accident and the normal-count states.

Finally, it is also worth mentioning that, in addition to negative
binomial models, we estimated Poisson models for the same accident
data and obtained similar results \citep[][]{M_08}.
In particular, we found that a two-state Markov switching Poisson (MSP)
model, which has the Poisson likelihood function instead of the NB
likelihood function in Eq.~(\ref{EQ_NB}), is strongly favored by
the empirical data as compared to standard zero-inflated Poisson models.

\section{Conclusions}
\label{CONCLUD}

A number of important observations can be made with regard to our
empirical findings. First, Markov switching count-data models provide
a superior statistical fit for accident frequencies relative to
standard zero-inflated models.
Second, Markov switching models, which explicitly consider transitions
between the zero-accident state and the normal-count state over time,
permit a direct empirical estimation of what states roadway segments
are in at different time periods. In particular, we found evidence
that some roadway segments changed their states over time (see
the bottom-right plot in Figure~\ref{FIGURE_2}).
Third, note that the Markov switching models avoid a theoretically
implausible assumption that some roadway segments are always zero-accident
because, in these models, every segment has a non-zero probability
of being in the normal-count state. Indeed, the long-term unconditional
mean of the accident rate for the $n^{\rm th}$ roadway segment is
equal to $\bar p^{(n)}_{1}\langle\lambda_{t,n}\rangle_t$, where
$\bar p^{(n)}_{1}=p^{(n)}_{0\rightarrow1}/(p^{(n)}_{0\rightarrow1}
+p^{(n)}_{1\rightarrow0})$ is the stationary probability of being in
the normal-count state $s_{t,n}=1$ and $\langle\lambda_{t,n}\rangle_t$ is
the time average of the accident rate in the normal-count state [refer to
Eq.~(\ref{EQ_LAMBDA})]. This long-term mean is always above zero
(see the bottom plot in Figure~\ref{FIGURE_3}), even for segments
that were likely to be in the zero-accident state over the whole
observed five-year time interval. Finally, we conclude that two-state
Markov switching count-data models are likely to be a better
alternative to zero-inflated models, in order to account for excess
of zeros observed in accident-frequency data.



\end{document}

%% file: ms_2_Table1a.tex
\begin{tabular}{|l|c|c|c|c|c|}
\hline
${}_{{}_{{}\atop{\displaystyle\mbox{\bf Variable}}}}$
 & \multicolumn{2}{|c|}{\bf\boldmath ZINB-$\tau^{\,\rm a}$} &
   \multicolumn{2}{|c|}{\bf\boldmath ZINB-$\gamma^{\,\rm b}$} & {\bf MSNB$^{\,\rm c}$}
\\
\cline{2-3}\cline{4-5}
 & {\bf by MLE} & {\bf by MCMC} & {\bf by MLE} & {\bf by MCMC}
 & {\bf by MCMC}
\\
\hline
\multicolumn{6}{|c|}{$\beta$- and $\alpha$-parameters in
Eq.~(\ref{EQ_LAMBDA})}\\
\hline
Intercept (constant term) &
$-15.0^{-12.5}_{-17.5}$ &
$-15.2^{-13.0}_{-17.4}$ &
$-11.6^{-8.32}_{-14.8}$ &
$-11.6^{-8.29}_{-14.6}$ &
$-17.3^{-13.0}_{-21.3}$\\
\hline
Accident occurring on interstates I-70 or I-164 (dummy)&
$-.683^{-.570}_{-.797}$ &
$-.685^{-.575}_{ -.794}$ &
$-.715^{-.602}_{-.829}$ &
$-.715^{-.593}_{-.836}$ &
$-.734^{-.617}_{-.850}$\\
\hline
Pavement quality index (PQI) average$^{\,\rm d}$ &
$-.0122^{-.0189}_{-.00550}$ &
$-.0122^{-.00562}_{-.0188}$ &
$-.0140^{-.00627}_{-.0217}$ &
$-.0143^{-.00643}_{-.0221}$ &
$-.0163^{-.00850}_{-.0240}$\\
\hline
Logarithm of road segment length (in miles) &
$.791^{.832}_{.751}$ &
$.791^{.829 }_{.754}$ &
$.929^{.978}_{.880}$ &
$.939^{.993}_{.886}$ &
$.887^{.929}_{.845}$\\
\hline
Number of ramps on the viewing side per lane per mile &
$.226^{.300}_{.153}$ &
$.227^{.306}_{.149}$ &
$.298^{.387}_{.209}$ &
$.304^{.394}_{.214}$ &
$.317^{.404}_{.230}$\\
\hline
Number of lanes on a roadway & -- &
-- &
-- &
-- &
$1.19^{2.04}_{.386}$\\
\hline
Median configuration is depressed (dummy) &
$.184^{.288}_{.0795}$ &
$.183^{.282}_{.0839}$ &
$.201^{.319}_{.0820}$ &
$.202^{.325}_{.0781}$ &
--\\
\hline
Median barrier presence (dummy) &
$-1.43^{-1.22}_{-1.64}$ &
$-1.43^{-1.14}_{-1.72}$ &
-- &
-- &
$-1.69^{-1.00}_{-2.46}$\\
\hline
Width of the interior shoulder is less that 5 feet (dummy) &
$.323^{.443}_{.202}$ &
$.323^{.434}_{.211}$ &
$.435^{.572}_{.297}$ &
$.437^{.569}_{.307}$ &
$.374^{.505}_{.243}$\\
\hline
Outside shoulder width (in feet) &
$-.0480^{-.0196}_{-.0764}$ &
$-.0478^{-.0207}_{-.0749}$ &
$-.0532^{-.0176}_{-.0887}$ &
$-.0532^{-.020}_{-.0867}$ &
$-.0537^{-.0214}_{-.0862}$\\
\hline
Outside barrier is absent (dummy) & -- &
-- &
$-.245^{-.117}_{-.373}$ &
$-.245^{-.101}_{-.389}$ &
$-.264^{-.124}_{-.403}$\\
\hline
Average annual daily traffic (AADT) &
$\displaystyle{-4.07^{-3.17\vphantom{O^o}}_{-4.97}}\atop\displaystyle{{}\times10^{-5}}$ &
$\displaystyle{-4.14^{-3.31\vphantom{O^o}}_{-5.04}}\atop\displaystyle{{}\times10^{-5}}$ &
$\displaystyle{-1.93^{-3.21\vphantom{O^o}}_{-6.50}}\atop\displaystyle{{}\times10^{-5}}$ &
$\displaystyle{-1.91^{-3.16\vphantom{O^o}}_{-5.83}}\atop\displaystyle{{}\times10^{-5}}$ &
$\displaystyle{-3.78^{-2.02\vphantom{O^o}}_{-5.26}}\atop\displaystyle{{}\times10^{-5}}$\\
\hline
Logarithm of average annual daily traffic &
$1.89^{2.17}_{1.61}$ &
$1.91^{2.16}_{1.67}$ &
$1.52^{1.88}_{1.15}$ &
$1.52^{1.86}_{1.15}$ &
$1.95^{2.34}_{1.49}$\\
\hline
Number of bridges per mile & -- &
-- &
-- &
-- &
$-.0214^{-.00164}_{-.0428}$\\
\hline
Maximum of reciprocal values of horizontal curve radii (in $1/{\rm mile}$) &
$-.140^{-.0710}_{-.209}$ &
$-.141^{-.0734}_{-.208}$ &
$-.134^{-.0559}_{-.213}$ &
$-.138^{-.0593}_{-.217}$ &
$-.106^{-.0289}_{-.183}$\\
\hline
Percentage of single unit trucks (daily average) &
$1.23^{1.84}_{.624}$ &
$1.23^{1.82}_{.646}$ &
$1.32^{1.96}_{.693}$ &
$1.32^{1.96}_{.691}$ &
$1.29^{1.90}_{.688}$\\
\hline
Number of changes per vertical profile along a roadway segment &
$.0555^{.0930}_{.0180}$ &
$.0562^{.0903}_{.0226}$ &
-- &
-- &
--\\
\hline
Over-dispersion parameter $\alpha$ in NB models &
$.144^{.183}_{.105}$ &
$.150^{.192}_{.114}$ &
$.130^{.168}_{.0925}$ &
$.142^{.185}_{.105}$ &
$.114^{.147}_{.0847}$\\
\hline
\end{tabular}

%% file: ms_2_Table1b.tex
\begin{tabular}{|l|c|c|c|c|c|}
\hline
${}_{{}_{{}\atop{\displaystyle\mbox{\bf Variable}}}}$
 & \multicolumn{2}{|c|}{\bf\boldmath ZINB-$\tau^{\,\rm a}$} &
   \multicolumn{2}{|c|}{\bf\boldmath ZINB-$\gamma^{\,\rm b}$} & {\bf MSNB}
\\
\cline{2-3}\cline{4-5}
 & {\bf by MLE} & {\bf by MCMC} & {\bf by MLE} & {\bf by MCMC}
 & {\bf by MCMC$^{\,\rm c}$}
\\
\hline
\multicolumn{6}{|c|}{$\tau$- and $\gamma$-parameters
in Eqs.~(\ref{EQ_Q_TAU}) and~(\ref{EQ_Q_GAMMA})}\\
\hline
The model parameter $\tau$ in Eq.~(\ref{EQ_Q_TAU}) &
$-1.72^{-1.45}_{-2.00}$ &
$-1.73^{-1.50}_{-1.98}$ &
-- &
-- &
--\\
\hline
Intercept (constant term) &
-- &
-- &
$23.1^{41.3}_{4.99}$ &
$26.5^{47.0}_{10.9}$ &
--\\
\hline
Logarithm of road segment length (in miles) &
-- &
-- &
$-1.34^{-.942}_{-1.73}$ &
$-1.4^{-1.03 }_{-1.83}$ &
--\\
\hline
Median barrier presence (dummy) &
-- &
-- &
$3.97^{4.86}_{3.08}$ &
$4.16^{5.20}_{3.27}$ &
--\\
\hline
Average annual daily traffic (AADT) &
-- &
-- &
$\displaystyle{9.23^{15.1\vphantom{O^o}}_{3.35}}\atop\displaystyle{{}\times10^{-5}}$ &
$\displaystyle{10.5^{17.4\vphantom{O^o}}_{5.72}}\atop\displaystyle{{}\times10^{-5}}$ &
--\\
\hline
Logarithm of average annual daily traffic &
-- &
-- &
$-2.88^{-.901}_{-4.86}$ &
$-3.28^{-1.59}_{-5.57}$ &
--\\
\hline
\hline
Mean accident rate ($\lambda_{t,n}$ for NB),
averaged over all values of ${\bf X}_{t,n}$ & -- &
$3.38$ &
-- &
$3.42$ &
$3.88$\\
\hline
Standard deviation of accident rate
($\sqrt{\lambda_{t,n}(1+\alpha\lambda_{t,n})}\,$ for NB), & & & & &\\
averaged over all values of explanatory variables ${\bf X}_{t,n}$ & -- &
$2.14$ &
-- &
$2.15$ &
$2.13$\\
\hline
Total number of free model parameters ($\beta$-s, $\gamma$-s, $\alpha$ and $\tau$) &
$16$ &
$16$ &
$19$ &
$19$ &
$16$\\
\hline
Posterior average of the log-likelihood (LL) & -- &
$-2510.68^{-2506.13}_{-2517.12}$ &
$--$ &
$-2436.34^{-2431.12}_{-2443.54}$ &
$-2124.82^{-2096.30}_{-2153.91}$\\
\hline
Max$(LL)$:\quad estimated max.~value of log-likelihood (LL) for MLE; & & & & &\\
maximum observed value of LL for Bayesian-MCMC &
$\displaystyle{-2502.67}\atop{\rm(MLE)}$ &
$\displaystyle{-2503.21}\atop{\rm(observed)}$ &
$\displaystyle{-2426.54}\atop{\rm(MLE)}$ &
$\displaystyle{-2427.41}\atop{\rm(observed)}$ &
$\displaystyle{-2049.45}\atop{\rm(observed)}$\\
\hline
Logarithm of marginal likelihood of data ($\ln[f({\bf Y}|{\cal M})]$) & -- &
$-2519.90^{-2516.95}_{-2521.59}$ &
-- &
$-2447.33^{-2443.93}_{-2448.86}$ &
$-2184.21^{-2186.70}_{-2169.56}$\\
\hline
Goodness-of-fit p-value & -- &
$0.005$ &
-- &
$0.177$ &
$0.191$\\
\hline
Maximum of the potential scale reduction factors (PSRF)$^{\,\rm e}$
& -- &
$1.01006$ &
-- &
$1.02200$ &
$1.02117$\\
\hline
Multivariate potential scale reduction factor (MPSRF)$^{\,\rm e}$
& -- &
$1.01023$ &
-- &
$1.02302$ &
$1.02189$\\
\hline
\multicolumn{6}{l}{$^{\rm a}$ Standard (conventional) \mbox{ZINB-$\tau$}
model estimated by maximum likelihood estimation (MLE) and Markov Chain
Monte Carlo (MCMC) simulations.}\\
\multicolumn{6}{l}{$^{\rm b}$ Standard \mbox{ZINB-$\gamma$}
model estimated by maximum likelihood estimation (MLE) and Markov Chain
Monte Carlo (MCMC) simulations.}\\
\multicolumn{6}{l}{$^{\rm c}$ Two-state Markov switching negative
binomial (MSNB) model where all reported parameters are for the normal-count state $s=1$.}\\
\multicolumn{6}{l}{$^{\rm d}$ The pavement quality index (PQI) is a composite
measure of overall pavement quality evaluated on a 0 to 100 scale.}\\
\multicolumn{6}{l}{$^{\rm e}$ PSRF/MPSRF are calculated separately/jointly
for all continuous model parameters. PSRF and MPSRF are close to 1 for
converged MCMC chains.}
\end{tabular}

%% file: ms_2_Table2.tex
\tabcolsep=0.45em
\renewcommand{\arraystretch}{1.45}
\begin{tabular}{|l|c|c|c|c|c|}
\hline
{\bf Variable} &
{\bf Mean} &
{\bf Standard deviation} &
{\bf Minimum} &
{\bf Median} &
{\bf Maximum}
\\
\hline
Accident occurring on interstates I-70 or I-164 (dummy) &
$.155$ &
$.363$ &
$0$ &
$0$ &
$1.00$
\\
\hline
Pavement quality index (PQI) average$^{\,\rm a}$ &
$88.6$ &
$5.96$ &
$69.0$ &
$90.3$ &
$98.5$
\\
\hline
Logarithm of road segment length (in miles) &
$-.901$ &
$1.22$ &
$-4.71$ &
$-1.03$ &
$2.44$
\\
\hline
Number of ramps on the viewing side per lane per mile &
$.138$ &
$.408$ &
$0$ &
$0$ &
$3.27$
\\
\hline
Number of lanes on a roadway &
$2.09$ &
$.286$ &
$2.00$ &
$2.00$ &
$3.00$
\\
\hline
Median configuration is depressed (dummy) &
$.630$ &
$.484$ &
$0$ &
$1.00$ &
$1.00$
\\
\hline
Median barrier presence (dummy) &
$.161$ &
$.368$ &
$0$ &
$0$ &
$1$
\\
\hline
Width of the interior shoulder is less that 5 feet (dummy) &
$.696$ &
$.461$ &
$0$ &
$1.00$ &
$1.00$
\\
\hline
Outside shoulder width (in feet) &
$11.3$ &
$1.74$ &
$6.20$ &
$11.2$ &
$21.8$
\\
\hline
Outside barrier absence (dummy) &
$.830$ &
$.376$ &
$0$ &
$1.00$ &
$1.00$
\\
\hline
Average annual daily traffic (AADT) &
$3.03\times10^4$ &
$2.89\times10^4$ &
$.944\times10^4$ &
$1.65\times10^4$ &
$14.3\times10^4$
\\
\hline
Logarithm of average annual daily traffic  &
$10.0$ &
$.623$ &
$9.15$ &
$9.71$ &
$11.9$
\\
\hline
Number of bridges per mile &
$1.76$ &
$8.14$ &
$0$ &
$0$ &
$124$
\\
\hline
Maximum of reciprocal values of horizontal curve radii (in $1/{\rm mile}$) &
$.650$ &
$.632$ &
$0$ &
$.589$ &
$2.26$
\\
\hline
Percentage of single unit trucks (daily average) &
$.0859$ &
$.0678$ &
$.00975$ &
$.0683$ &
$.322$
\\
\hline
Number of changes per vertical profile along a roadway segment &
$.522$ &
$.908$ &
$0$ &
$0$ &
$6.00$
\\
\hline
\multicolumn{6}{l}{$^{\rm a}$ The pavement quality index (PQI) is a composite
measure of overall pavement quality evaluated on a 0 to 100 scale.}
\end{tabular}

%% file: ms_2.bbl
\begin{thebibliography}{}

\bibitem[Bohning et al.(1999)]{BDSMK_99}
Bohning, D., Dietz, E., Schlattmann, P., Mendonca L., Kirchner, U.,
1999: The zero-inflated Poisson model and the decayed, missing
and filled teeth index in dental epidemiology. Journal of Royal
Statistical Society A 162(2), 195–209.

\bibitem[Broek(1995)]{B_95}
van den Broek, J., 1995: A score test for zero-inflation in a
Poisson distribution. Biometrics 51(2), 738–743

\bibitem[Carson and Mannering(2001)]{CM_01}
Carson, J., Mannering, F. L., 2001. The effect of ice warning signs
on ice-accident frequencies and severities. Accid. Anal. Prev.
33(1), 99-109.

\bibitem[Cowan(1998)]{C_98}
Cowan, G., 1998. Statistical Data Analysis. Clarendon Press,
Oxford Univ. Press, USA

\bibitem[Kass and Raftery(1995)]{KR_95}
Kass, R. E., Raftery, A. E., 1995. Bayes Factors.
J. Americ. Statist. Assoc. 90(430), 773-795.

\bibitem[Lambert(1992)]{L_92}
Lambert, D., 1992: Zero-inflated Poisson regression, with an
application to defects in manufacturing. Technometrics 34(1), 1–14.

\bibitem[Lee and Mannering(2002)]{LM_02}
Lee, J., Mannering, F. L., 2002. Impact of roadside features on the
frequency and severity of run-off-roadway accidents: an empirical
analysis. Accid. Anal. Prev. 34(2), 149-161.

\bibitem[Lord et al.(2005)]{LWI_05}
Lord, D., Washington, S., Ivan, J. N., 2005. Poisson, Poisson-gamma and
zero-inflated regression models of motor vehicle crashes: balancing
statistical fit and theory. Accid. Anal. Prev. 37(1), 35-46.

\bibitem[Lord et al.(2007)]{LWI_07}
Lord, D., Washington, S., Ivan, J. N., 2007. Further notes on the
application of zero-inflated models in highway safety.
Accid. Anal. Prev. 39(1), 53-57.

\bibitem[Maher and Summersgill(1996)]{MS_96}
Maher M. J., Summersgill, I., 1996. A comprehensive methodology for
the fitting of predictive accident models.
Accid. Anal. Prev. 28(3), 281-296.

\bibitem[Malyshkina(2006)]{M_06}
Malyshkina, N. V., 2006. Influence of speed limit on roadway safety in
Indiana. MS thesis, Purdue University.
http://arxiv.org/abs/0803.3436

\bibitem[Malyshkina(2008)]{M_08}
Malyshkina, N. V., 2008. Markov switching models: an application to
roadway safety. PhD thesis in preparation, Purdue University.
http://arxiv.org/abs/0808.1448

\bibitem[Malyshkina et al.(2009)]{MMT_09}
Malyshkina, N. V., Mannering, F. L., Tarko, A. P., 2009.
Markov switching negative binomial models: an application to
vehicle accident frequencies. Accid. Anal. Prev. 41(2), 217-226.

\bibitem[Miaou(1994)]{M_94}
Miaou, S. P., 1994. The relationship between truck accidents and
geometric design of road sections: Poisson versus negative binomial
regressions. Accid. Anal. Prev. 26(4), 471-482.

\bibitem[Press et al.(2007)]{PTVF_07}
Press, W. H., Teukolsky, S. A., Vetterling, W. T., Flannery B. P., 2007.
Numerical Recipes 3rd Edition: The Art of Scientific Computing.
Cambridge Univ. Press, UK.

\bibitem[Robert(2001)]{R_01}
Robert, C. P., 2001. The Bayesian choice: from decision-theoretic
foundations to computational implementation.
Springer-Verlag, New York.

\bibitem[Shankar et al.(1997)]{SMM_97}
Shankar, V., Milton, J., Mannering, F., 1997. Modeling accident
frequencies as zero-altered probability processes: an empirical
inquiry. Accid. Anal. Prev. 29(6), 829-837.

\bibitem[Spiegelhalter et al.(2002)]{SBCL_02}
Spiegelhalter, D. J., Best, N. G., Carlin, B. P.,
van der Linde, A., 2002. Bayesian measures of model complexity
and fit. J. Royal Stat. Soc. B, {\bf 64}, 583-639.

\bibitem[Tsay(2002)]{T_02}
Tsay, R. S., 2002. Analysis of financial time series: financial
econometrics. John Wiley \& Sons, Inc.

\bibitem[Washington et al.(2003)]{WKM_03}
Washington, S. P., Karlaftis, M. G., Mannering, F. L., 2003. Statistical
and econometric methods for transportation data analysis.
Chapman \& Hall/CRC.

\bibitem[Wood(2002)]{W_02}
Wood, G. R., 2002. Generalised linear accident models and
goodness of fit testing. Accid. Anal. Prev. 34, 417-427.

\end{thebibliography}
